# Potential jumps at transport bottlenecks cause instability of nominally ionic solid electrolytes in electrochemical cells

Yanhao Dong[1], Zhichao Zhang[2], Ana Alvarez[2], I-Wei Chen[2]

[1]Department of Nuclear Science and Engineering, Massachusetts Institute of Technology, Cambridge, MA 02139, USA

[2]Department of Material Science and Engineering, University of Pennsylvania, Philadelphia, PA 19104, USA

## Abstract

Normal operations of electrochemical devices such as solid oxide fuel cells (SOFC), solid oxide electrolyzer cells (SOEC) and lithium ion batteries (LIB) sometimes fail because of unexpected formation of internal phases. These phases include oxygen bubbles at grain boundaries inside the zirconia electrolyte of SOEC, isolated Li metal islands inside the (garnet type) $Li_7La_3Zr_2O_{12}$ electrolyte of all-solid-state LIB, and similar Na metal islands inside the Na-beta-alumina and NASICON electrolytes of Na-S batteries. Remarkably, although the devices can operate in both polarities, the propensity for failure depends on the polarity. Here we explain these and other phenomena in nominally ionic solid electrolytes and mixed-conducting electrodes in simple thermodynamic and kinetic terms: the unexpected internal phases are caused by a large potential jump that is needed to push a constant ion or electron flow through its internal transport bottleneck. Definite rules for internal phase formation including its polarity dependence are formulated to help predict and mitigate it, which leads to microstructural instability, efficiency deterioration and breakdown.

**Keywords:** SOFC; Li-ion battery; Superionic conductor; Degradation; Grain boundaries

### I. Introduction

Many electrochemical devices comprising a solid electrolyte and two electrodes can operate reversibly. For example, solid oxide fuel cells (SOFC) [1-4] convert chemical energy to electricity, while under reverse mode they become solid oxide electrolyzer cells (SOEC) [4-10] generating fuel from electricity. Reversible operation is even routine in rechargeable all-solid-state lithium ion batteries (LIB) [11-13], which can sustain many cycles of charge and discharge. However, the damages accrued in their services are often polarity dependent, as evidenced by a more severe degradation in SOEC than in SOFC, and more in LIB charging than in LIB discharging. In principle, these devices are supposed to operate within a safe thermodynamic window—specified by the electrode potentials—and no phase instability inside the solid electrolyte is expected. Yet oxygen bubbles have been found in SOEC (not in SOFC) at the grain boundaries inside the zirconia electrolyte away from the oxygen electrode [4, 7-10], which implicates an internal oxygen pressure higher than that of the oxygen electrode. Likewise, isolated Li metal islands (not in the form of dendrites) form inside the $Li_7La_3Zr_2O_{12}$ electrolyte away from the Li metal electrode in an all-solid-state LIB [14-16], and similar Na metal islands form inside the Na-beta-alumina [17-19] or NASICON [19] solid electrolytes in Na-S batteries. Clearly, the thermodynamic design has failed, and the internal chemical potentials of $O_2$, Li and Na—all charge-neutral molecular or atomic species—have exceeded their limiting potentials under normal operation conditions.

These observations have been variously explained in terms of stress/field/current concentrations [20, 21], electronic short circuit [21], and electrode overpotential [22-24], which is the excess potential required to drive sluggish interfacial reactions at electrodes. The latter thinking is: once the overpotential is included, which can be quite large, the electrode potentials may have exceeded the thermodynamic threshold. In the "consensus" view, the thermodynamic potentials inside the device are still bounded by these modified boundary values, and they cannot be worse than what is caused by an overpotential. For example, for SOFC and SOEC, it is thought "oxygen pressure inside the electrolyte will never become higher than the pressure corresponding to the electrode potential of the oxygen electrode and never lower than corresponding to the electrode potential of the hydrogen electrode, irrespective of which mode or condition for the cell operation." [22] We will challenge such consensus, however, because we have found, at transport bottlenecks of ion or electron flow inside the solid electrolytes, potential jumps that are so large as to place the chemical potential of a charge-neutral species beyond its boundary values at two electrodes. Just like electrode overpotentials, these potential jumps have a similarly adverse effect; they can engender unexpected *neutral* phases to form internally, which almost always leads to stresses, distortions, fractures and short-circuits resulting in device deterioration and failures. [4, 7-10, 14-19] Importantly, the magnitude of these jumps strongly depends on the polarity of the operation. With these features, our theory can now provide a unified, self-consistent explanation of all the internal damages mentioned above. As will be described later, it can also explain other microstructure and device-specific observations that are relevant to device failures.

Below we will first identify the origins of transport-induced potential jumps in **Section II**. Next, in **Section III**, we demonstrate how the jumps manifest themselves in polycrystalline solid electrolytes causing failures under practically relevant SOEC/SOFC conditions. In **Section IV**, the pivotal importance of seemingly minuscule electronic conductivity is elucidated, which leads to a new suggestion for damage mitigation: by trace-element doping. Practical predictions based on the above findings are summarized in **Section V**, which is followed by discussions on experimental observations in **Section VI** before concluding in **Section VII**. We will use the conductivity data of yttria-stabilized cubic zirconia (abbreviated as YSZ hereafter) for numerical calculations following the method described in **Appendix 1** and **2**, and more specific analyses of batteries are described in **Appendix 3-4**. Additional justifications for the method and assumptions are given on-line in **Appendix 5-7** of Ref. [25].

## II. Origin of transport-induced potential jump

2.1 Transport bottlenecks in ionic and electronic channels

Transport bottlenecks in solid electrolytes may exist in both ionic and electronic channels. In YSZ, a fast $O^{2-}$ conductor, grain boundaries are bottlenecks in the ionic channel because their $O^{2-}$ conductivity $\sigma_{O^{2-}}$ at ~800°C is only 1/100th of that of the lattice [26, 27]. Transport bottlenecks in the electronic channel can likewise arise from compositional or structural inhomogeneities, but the most important bottleneck is a generic one located at the minimum of $\sigma_e + \sigma_h$, the sum of electron and hole conductivities, as shown in **Fig. 1a**. Such minimum exists in every solid electrolyte that undergoes severe redox reactions [28-30], and as alluded to by Wagner in 1933 [31] and later quantitatively demonstrated by others [22, 32, 33], at this minimum the oxygen chemical potential $\mu_{O_2}$ must acquire a steep gradient at the steady state. Indeed, for YSZ under the steady-state operating condition appropriate for SOEC, **Fig. 1b** shows an almost step-like "jump" in $\mu_{O_2}$ at precisely $(\sigma_e + \sigma_h)_{min}$, at $x/L$=0.57, marked by a vertical dash line. Below, we will find similar $\mu_{O_2}$ jumps at other transport bottlenecks. This is true even when the bottleneck is in a seemingly unimportant channel. For example, under normal operation conditions, the electronic channel in YSZ featuring a large $\mu_{O_2}$ jump in **Fig. 1b** has only a miniscule $\sigma_e + \sigma_h$ everywhere compared to the ionic channel that has a large and $\sigma_{O^{2-}}$.

Though the jump in **Fig. 1b** is not unique, it is especially important because it places a large part of the electrolyte under a $\mu_{O_2}$ very close to those of the electrodes. In effect, at the steady state it renders the oxygen-electrode side of the electrolyte a virtual oxygen electrode and the hydrogen-electrode side a virtual hydrogen electrode, thus subjecting them to almost the full influence of the electrode's overpotentials, which quite often reflect the most extreme electrochemical conditions including polarization. Because of this, a large part of the electrolyte is prone to damage: if there is a further $\mu_{O_2}$ jump at another bottleneck, then the local $\mu_{O_2}$ may well exceed the boundary $\mu_{O_2}$ of the oxygen-electrode and cause O$_2$ bubbles to form inside the electrolyte [4, 7-10]. The opposite may also happen: at another bottleneck $\mu_{O_2}$ may well fall below the boundary $\mu_{O_2}$ of the hydrogen-electrode and cause internal voids to form [34, 35]. These possibilities are schematically depicted in **Fig. 2**.

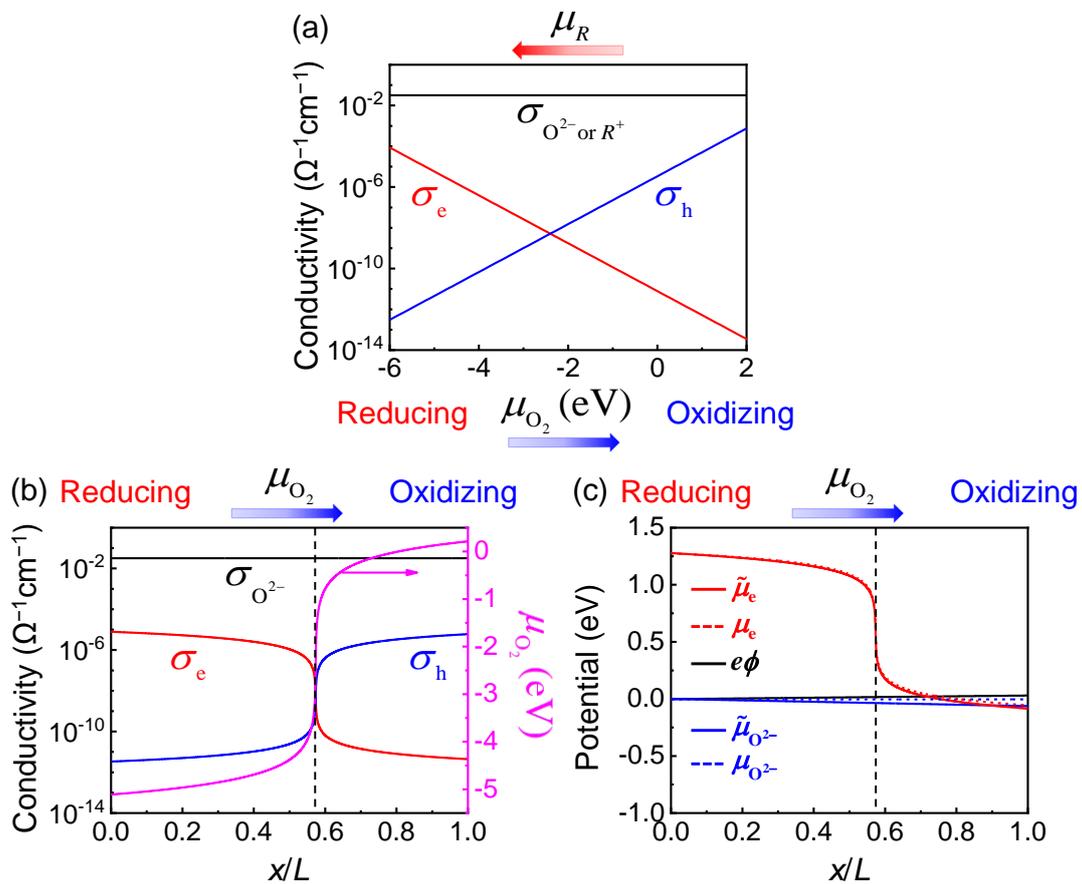

**Figure 1 Conductivities and potential distributions in "single-crystal" solid electrolytes at 800°C.** (a) Redox-insensitive ionic conductivity overwhelms redox-dependent electronic conductivities in YSZ at all practical $\mu_{O_2}$ [28]. Same

trend holds for other fast $O^{2-}$ [29, 30] and fast-$R^+$ ($R$=Li, Na) [37-39] solid electrolytes at all practical $\mu_{O_2/R}$. Distributions of (b) $\mu_{O_2}$ and conductivities, and (c) electrostatic potential $\phi$, electrochemical potentials and chemical potentials of $O^{2-}$ and electrons in YSZ SOEC at $-1$ A/cm$^2$. ($\tilde{\mu}_{O^{2-}} = \mu_{O^{2-}} - 2e\phi$, $\tilde{\mu}_e = \mu_e - e\phi$, and $\mu_{O_2} = \mu_{O_2}^{\Theta} + kT \ln PO_2$ where $\mu_{O_2}^{\Theta} \equiv 0 =$ oxygen chemical potential at standard condition of 1 atm.) Other conditions: $\mu'_{O_2} = -5.11$ eV ($PO_2$=10$^{-24}$ atm) at $x$=0, $\mu''_{O_2} = 0.21$ eV ($PO_2$=10 atm) at $x$=$L$, electrolyte thickness $L$=10 μm. Calculated ionic vs. electronic contribution to total current density is 0.9986 vs. 0.0014.

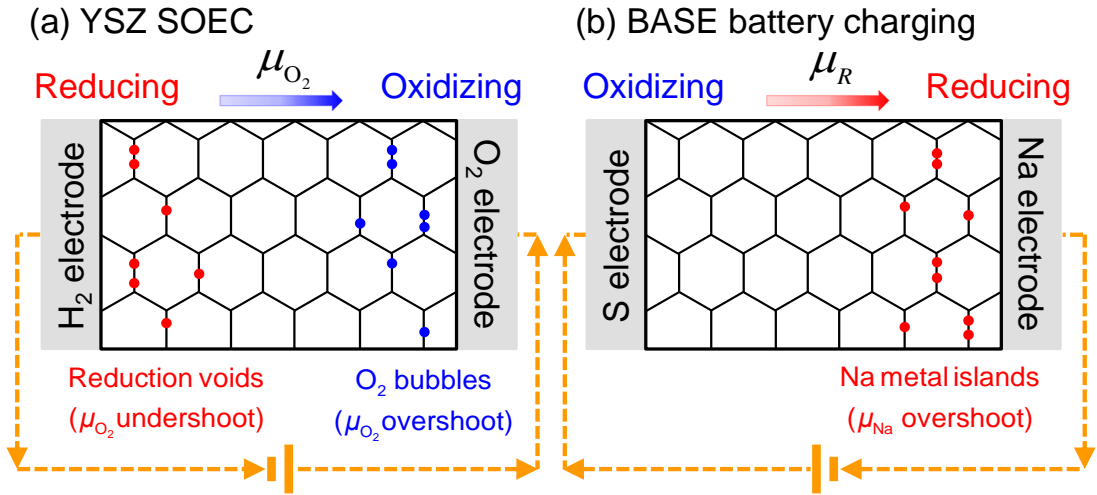

**Figure 2 Exemplary unexpected phase formation at transport bottlenecks**. Left: oxygen gas bubbles and reduction voids at low-$\sigma_{O^{2-}}$ grain boundaries; right: sodium metal islands at low-$\sigma_{Na^+}$ grain boundaries. Phases preferentially form on transverse grain boundaries because they carry the most ion flow, experience the most potential overshoot/undershoot and sustain the most driving force for phase formation.

2.2 $\mu_{O_2}$ jumps at transport bottlenecks in YSZ

Why is the chemical potential of $O_2$, a neutral molecular species not supposed to permeate through the device at all, correlated to the transport bottlenecks of any charged species? The chemical potential of electron $\mu_e$ plotted in **Fig. 1c** offers a clue. At $x/L=0.57$, $\mu_e$ has a step-like drop, and except for a factor of 1/4, the variation of $\mu_e$ is exactly the mirror image of that of $\mu_{O_2}$, which has a step-like jump in **Fig. 1b**. We will refer to this correlation as $\delta\mu_{O_2} = -4\delta\mu_e$. It suggests that the two variations are linked by a reaction, which turns out to be the following redox reaction

$$O^{2-} = \tfrac{1}{2}O_2 + 2e \qquad (1)$$

This is because if we assume local equilibrium (more on this later), then Eq. (1) demands

$$\mu_{O^{2-}} = \tfrac{1}{2}\mu_{O_2} + 2\mu_e \qquad (2)$$

Since YSZ having a fixed concentration of $O^{2-}$ (and oxygen vacancies $V_O^{\bullet\bullet}$) has the same $\mu_{O^{2-}}$—the chemical potential of $O^{2-}$—everywhere, it sets the left hand side of Eq. (2) constant. Therefore, the right-hand side is also constant and must satisfy $\delta\mu_{O_2} = -4\delta\mu_e$. Incidentally, the $\mu_{O_2}$ dependence of conductivity depicted in **Fig. 1a** also comes from Eq. (1), because via the relationship between log(concentration) and chemical potential, we can obtain the electron concentration proportional to $\exp(-\mu_{O_2}/4k_BT)$, where $k_B$ is the Boltzmann constant and $T$ is the absolute temperature. Moreover, since the hole concentration is inversely proportional to that of electron, it is proportional to $\exp(\mu_{O_2}/4k_BT)$ and we can obtain hole's chemical potential as $-\mu_e$.

In YSZ, $O^{2-}$ has no gradient in chemical potential, so its flow is driven by the gradient in electrostatic potential $\phi$ only. However, electron has a gradient in both the chemical potential and $\phi$, so its flow is driven by the gradient of the electrochemical potential. We consider two cases. (i) An electronic bottleneck (e.g., the one depicted in **Fig. 1** at $(\sigma_e + \sigma_h)_{min}$): Excluding the possibility that it is an internal crack, there is no reason why this bottleneck must coincide with an ionic bottleneck. So, at the steady state it should support a constant $O^{2-}$ flow, hence there is no jump in $\phi$ here. This is verified in **Fig. 1c**, where the only jump is in $\mu_e$, which is required to push a constant electron flow through the electronic bottleneck. This $\mu_e$ jump necessitates a $\mu_{O_2}$ jump because of $\delta\mu_{O_2} = -4\delta\mu_e$ as mentioned before. (ii) An ionic bottleneck: Here there is a need for a jump in $\phi$ because it provides the only driving force that pushes a constant $O^{2-}$ flow. Like before, there is no reason why the ionic bottleneck must coincide with an electronic bottleneck. So, at the steady state, there should be a constant electronic flow and no jump in the electrochemical potential of electron, $\tilde{\mu}_e = \mu_e - e\phi$. But to keep the electrochemical potential constant, the $\phi$ jump must be canceled out by a $\mu_e$ jump, which in turn translates to a $\mu_{O_2}$ jump due to $\delta\mu_{O_2} = -4\delta\mu_e$ as before.

In summary, in YSZ, we have $\delta\mu_{O_2} = -4\delta\mu_e$ everywhere, a very large $\delta\mu_{O_2} = -4\delta\mu_e$ at any electronic bottleneck, and $\delta\mu_{O_2} = -4e\delta\phi$ at any ionic bottleneck. Together, they demand a $\mu_{O_2}$ jump at every transport bottleneck whichever channel it lies in.

## 2.3 $\mu_R$ jumps at transport bottlenecks in Li and Na batteries (R=Li, Na)

We can generalize the above concept to batteries that have "good" solid electrolytes, such as $Li^+$-conducting $Li_7La_3Zr_2O_{12}$ (LLZO) and $Na^+$-conducting beta-alumina solid electrolyte (BASE) and NASICON. In these electrolytes, the concentration of the fast ion $R^+$ (R=Li or Na) hence its conductivity $\sigma_{R^+}$ and chemical potential $\mu_{R^+}$, are insensitive to the redox conditions of

$$R = R^+ + e \quad (3)$$

which demands

$$\mu_R = \mu_{R^+} + \mu_e \quad (4)$$

Indeed, if we use hole—the antiparticle of electron—to rewrite Eq. (3) as $R^+ = R + h$ and Eq. (4) as $\mu_{R^+} = \mu_R + \mu_h$, then they formally correspond to Eq. (1) and (2), respectively. Therefore, from Eq. (3-4), very similar results follow as discussed above. With a constant $\mu_{R^+}$, we have $\delta\mu_R = \delta\mu_e$ in general, $\delta\mu_R = e\delta\phi$ at the ionic bottleneck, an electron concentration proportional to $\exp(\mu_R/k_BT)$ and a hole concentration proportional to $\exp(-\mu_R/k_BT)$. Note that the reducing condition is met at high Li/Na activities that is equivalent to low $O_2$ activities, and the oxidizing condition met at low Li/Na activities that is equivalent to high $O_2$ activities. This correspondence allows us to use the same **Fig. 1-2** to show analogous redox behavior, as directed by the trend-directing arrows of $\mu_{O_2}$ and $\mu_R$.

## 2.4 Local equilibrium and non-equilibrium

Before proceeding, we digress to make three comments. First, we have followed standard electrochemistry [22-24, 36] to assume local equilibrium, which is the basis of redox reactions Eq. (1-4). (For solid electrolytes under the normal operational conditions of electrochemical cells, this can be explicitly justified by thermodynamic and kinetic calculations of the transient lengths and time scales as shown on-line in **Appendix 5** and **6** of Ref. [25].) Second, because in a good electrolyte like YSZ $O^{2-}$ is unlikely to change valence, the $O^{2-}$ flow and the electron-hole flow cannot crosstalk and must each remain constant at the steady state, or else $O^{2-}$ or e/h will accumulate/deplete internally. This also applies to $Na^+$ and $Li^+$ in a good electrolyte. Third, internal potential jumps explored in this work has a different characteristic from electrode overpotentials. Overpotentials develop under far-from-equilibrium conditions to drive local interfacial reactions, which are non-linear reactions that become grossly irreversible when overpotentials are large. In contrast, our potential jumps develops under local equilibrium conditions to aid local diffusion, which still follows the linear response theory even though the potential jumps may become large enough to form new phases.

**III. Numerical results on polycrystalline electrolytes under practical conditions**

3.1 $\mu_{O_2}$ can overshoot/undershoot boundary electrode potentials

The $\mu_{O_2}$ distribution coming from the $(\sigma_e + \sigma_h)_{min}$ illustrated in **Fig. 1b-c** was for a YSZ without any grain boundary, i.e., it is in a YSZ single crystal. We next use a model polycrystal containing several grain boundaries as ionic bottlenecks to illustrate the impact of additional $\mu_{O_2}$ jumps at these bottlenecks. We compute all the potentials inside a YSZ in SOEC (**Fig. 3**) and SOFC (**Fig. 4**) at 800°C using the formulation in **Appendix 1**, which closely follows that in Ref. [33]. Here, O$^{2-}$ flows from the hydrogen electrode (at $x=0$) to the oxygen electrode (at $x=L$, where $L$ is the YSZ thickness) in SOEC, and in the opposite direction in SOFC. (See **Appendix 2** on how the flow directions are determined.)

The model polycrystal has three grain boundaries equally spaced at $\Delta x = \frac{1}{4}L$ apart. In **Fig. 3a**, $\sigma_{O^{2-}}$ of the last boundary ($x=\frac{3}{4}L$) $\sigma_{O^{2-}}^{\text{Last GB}}$ is set at $10^{-2}$, $10^{-3}$, or $10^{-4}$ of the lattice conductivity $\sigma_{O^{2-}}^{L}$, while the other two boundaries have $10^{-2}\sigma_{O^{2-}}^{L}$; the same applies to **Fig. 3b** but it is the first boundary ($x=\frac{1}{4}L$) that is assigned a different $\sigma_{O^{2-}}^{\text{First GB}}$. The middle panels of **Fig. 3a-b** illustrates how $\sigma_{O^{2-}}$, $\sigma_e$ and $\sigma_h$ are distributed for a polycrystal with one boundary having $10^{-4}\sigma_{O^{2-}}^{L}$. Under operating conditions (boundary oxygen potentials and current density) of a YSZ SOEC similar to those in Ref. [4], we see the $\mu_{O_2}$ jump at the last boundary in **Fig. 3a** becomes more severe as $\sigma_{O^{2-}}^{\text{Last GB}}$ decreases; $\mu_{O_2}$ actually exceeds the boundary potential at $x=L$ when $\sigma_{O^{2-}}^{\text{Last GB}}/\sigma_{O^{2-}}^{L} = 10^{-4}$. Likewise, in **Fig. 3b**, we see a $\mu_{O_2}$ jump at the first boundary causes $\mu_{O_2}$ to fall below the boundary potential at $x=0$ when $\sigma_{O^{2-}}^{\text{First GB}}/\sigma_{O^{2-}}^{L} = 10^{-4}$. Inspection of these figures makes it clear why the jumps at the grain boundaries exceed the boundary potential: a major reason is because they are built on a very large jump at $(\sigma_e + \sigma_h)_{\min}$, which is an electronic bottleneck. This confirms what we said in **Section 2.1**, that the large jump at $(\sigma_e + \sigma_h)_{\min}$ renders the oxygen-electrode side of the electrolyte a virtual oxygen electrode and the hydrogen-electrode side a virtual hydrogen electrode, which makes both sides susceptible to $\mu_{O_2}$ overshoots/undershoots. Inspection of $e\phi$, $\mu_e$, $\mu_{O^{2-}}$ and the corresponding electrochemical potentials ($\tilde{\mu}_e$ and $\tilde{\mu}_{O^{2-}}$) in the lower panels of **Fig. 3a-b** further confirms $\delta\mu_{O_2} = -4\delta\mu_e$ at every $\mu_{O_2}$ jump and $\delta\mu_{O_2} = -4e\delta\phi$ at every grain-boundary (ionic bottleneck) jump; this is exactly what we stated at the end of **Section 2.2**. Very similar features are seen in a YSZ SOFC (**Fig. 4**) except for one crucial difference: there is no $\mu_{O_2}$ overshoot/undershoot beyond the boundary potentials at *x*=0 or *L* anywhere at all.

These results demonstrate that, (i) notwithstanding electron's minuscule contribution (0.14% to 0.29%) to the total current (as stated in the caption of **Fig. 3** and **Fig. 4**), the redox-reaction (Eq. (1)) is crucial to the variation of $\mu_{O_2}$ and the electronic bottleneck at $(\sigma_e + \sigma_h)_{min}$ is pivotal to the electrolyte stability in a nominally ionic solid electrolyte [22-24, 33], (ii) the key features of rapid changes in the potential profiles at transport bottlenecks can be accurately predicted by "$\delta\mu_{O_2} = -4\delta\mu_e$ everywhere including at any electronic bottleneck, and $\delta\mu_{O_2} = -4e\delta\phi$ at any ionic bottleneck," and (iii) even though the YSZ device can be operated under either polarity, the likelihood of potential overshoot/undershoot, hence damage, is polarity dependent and much stronger in SOEC,

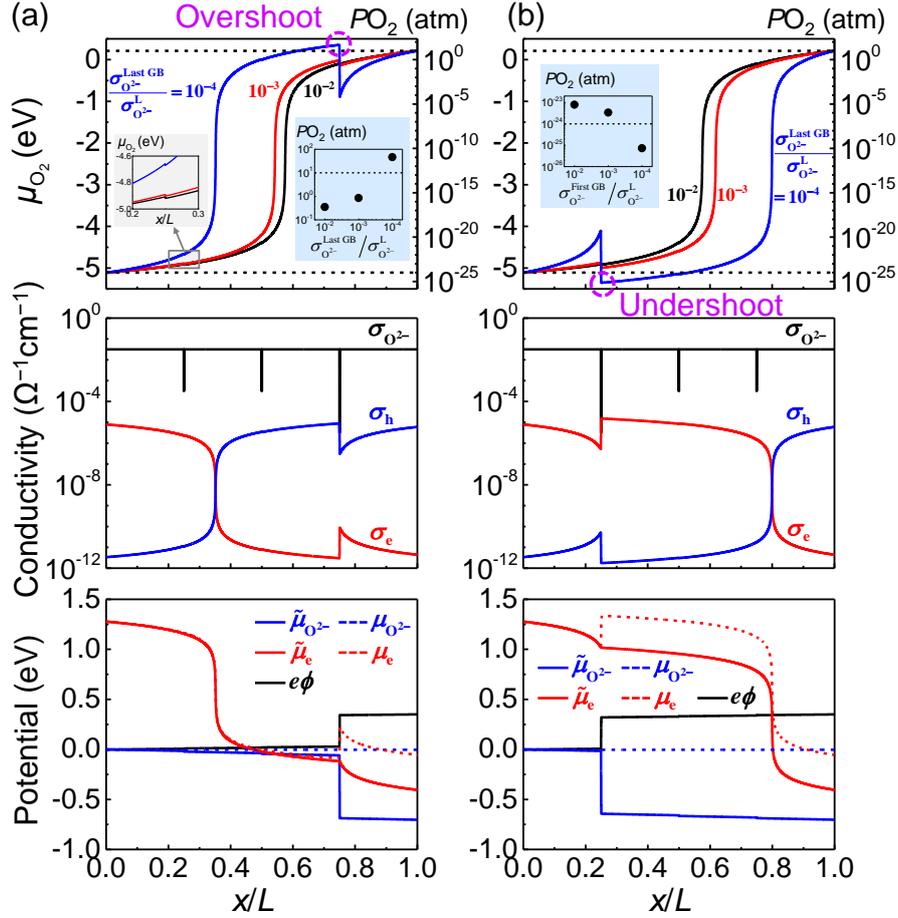

**Figure 3 Calculated potentials and conductivities inside "polycrystalline" solid electrolytes in SOEC.** YSZ electrolyte in SOEC at −1 A/cm$^2$ with grain boundaries at ¼$L$, ½$L$ and ¾$L$ from left and (a) $\sigma_{O^{2-}}^{\text{Last GB}}/\sigma_{O^{2-}}^{L}$ and (b) $\sigma_{O^{2-}}^{\text{First GB}}/\sigma_{O^{2-}}^{L}$ set as $10^{-2}$, $10^{-3}$ or $10^{-4}$, while $\sigma_{O^{2-}}$ of all other boundaries at $10^{-2}\sigma_{O^{2-}}^{L}$. Upper panels: $\mu_{O_2}$ vs. $x/L$ with dashed lines indicating boundary $\mu_{O_2}$ at two electrodes. Circles in violet indicate $\mu_{O_2}$ has exceeded boundary values, thus termed overshoot/undershoot. Insets with blue shadow show unbounded overshoot/undershoot is mathematically possible. Inset with grey shadow shows magnified view of $\mu_{O_2}$ at $x/L$=¼. Middle panels: conductivities of O$^{2-}$, electrons and holes vs. $x/L$ for $\sigma_{O^{2-}}^{\text{Last/first GB}}/\sigma_{O^{2-}}^{L} = 10^{-4}$. Lower panels: electrostatic potential $\phi$, electrochemical potentials and chemical potentials of O$^{2-}$ and e, for $\sigma_{O^{2-}}^{\text{Last/first GB}}/\sigma_{O^{2-}}^{L} = 10^{-4}$. Other conditions: 800°C with

$\mu'_{O_2} = -5.11$ eV ($PO_2=10^{-24}$ atm) at $x=0$, $\mu''_{O_2} = 0.21$ eV ($PO_2=10$ atm) at $x=L$, $L=10$ μm, grain boundary thickness $t=10$ nm. Calculated ionic vs. electronic contribution to total current density in upper panels are (a) 0.9986 vs. 0.0014 for black curve, 0.9985 vs. 0.0015 for red curve and 0.9978 vs. 0.0022 for blue curve; (b) 0.9986 vs. 0.0014 for black curve, 0.9984 vs. 0.0016 for red curve and 0.9971 vs. 0.0029 for blue curve.

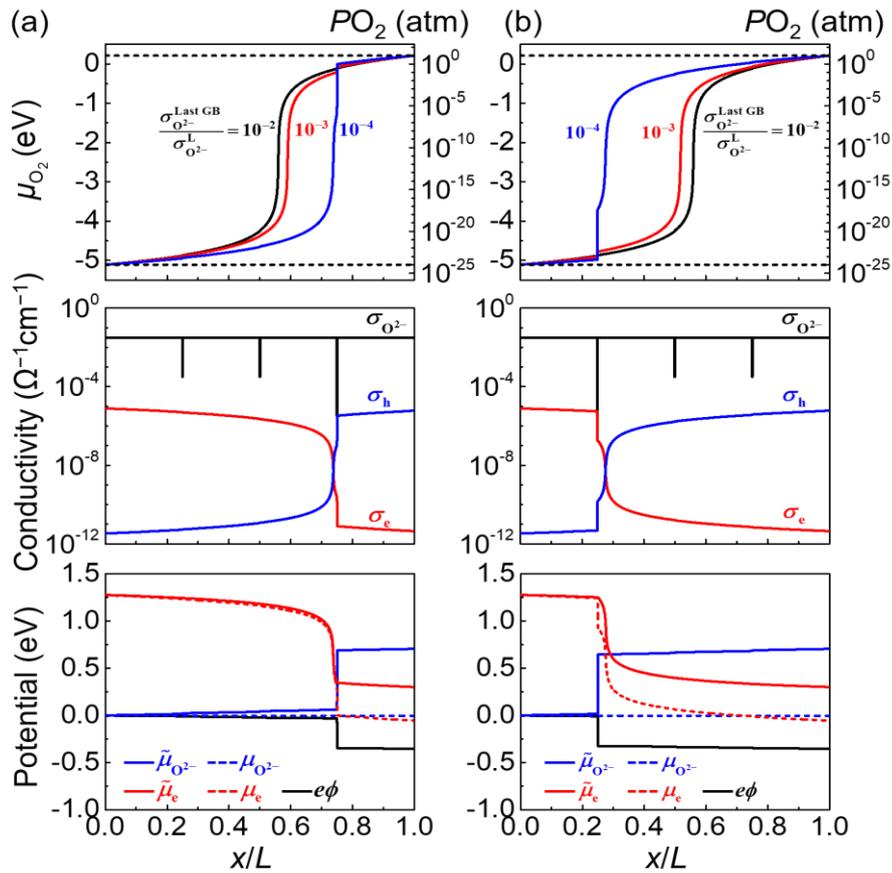

**Figure 4 Calculated potentials and conductivities inside "polycrystalline" solid electrolytes in SOFC.** YSZ electrolyte in SOFC at 1 A/cm² with grain boundaries at ¼L, ½L and ¾L from left and (a) $\sigma_{O^{2-}}^{\text{Last GB}}/\sigma_{O^{2-}}^{L}$ and (b) $\sigma_{O^{2-}}^{\text{First GB}}/\sigma_{O^{2-}}^{L}$ set as $10^{-2}$, $10^{-3}$, or $10^{-4}$, while $\sigma_{O^{2-}}$ of all other boundaries at $10^{-2}\sigma_{O^{2-}}^{L}$. Upper panels: $\mu_{O_2}$ vs. $x/L$ with dashed lines indicating boundary $\mu_{O_2}$ at two electrodes. Middle panels:

conductivities of $O^{2-}$, electrons and holes, for $\sigma_{O^{2-}}^{\text{Last/first GB}}/\sigma_{O^{2-}}^{\text{L}}=10^{-4}$. Lower panels: electrostatic potential $\phi$, electrochemical potentials and chemical potentials of $O^{2-}$ and e, for $\sigma_{O^{2-}}^{\text{Last/first GB}}/\sigma_{O^{2-}}^{\text{L}}=10^{-4}$. Other conditions: 800°C with $\mu'_{O_2}=-5.11$ eV ($PO_2=10^{-24}$ atm) at $x=0$, $\mu''_{O_2}=0.21$ eV ($PO_2=10$ atm) at $x=L$, $L=10$ μm, $t=10$ nm. Ionic vs. electronic contributions to total current density in upper panels are (a) 1.0012 vs. 0.0012 for black curve, 1.0011 vs. 0.0011 for red curve and 1.0008 vs. 0.0008 for blue curve, and (b) 1.0012 vs. 0.0012 for black curve, 1.0010 vs. 0.0010 for red curve and 1.0008 vs. 0.0008 for blue curve.

3.2 Why more damage in YSZ SOEC and during battery charging

Why in the exact, numerical results in **Fig. 3-4** is there a $\mu_{O_2}$ overshoot/undershoot in SOEC but not in SOFC—the two differing only in the direction of $O^{2-}$ flow? This can be more intuitively explained using the schematics of **Fig. 5e-f** in which the nonlinear base-line $\mu_{O_2}$ (e.g. the ones in **Fig. 1b** and **Fig. 3**) is simplified to a straight line between the boundary $\mu_{O_2}$ of the two electrodes. As already mentioned, in the case of an ionic bottleneck, there must be a $\phi$ jump to drive the $O^{2-}$ flow in the correct direction, which is leftward in SOFC (**Fig. 5a**) and rightward in SOEC (**Fig. 5b**). By way of $\delta\mu_{O_2}=-4e\delta\phi$, this also determines the direction of the $\mu_{O_2}$ jump (rightward in SOFC in the upper panel of **Fig. 5e** and leftward in SOEC in the upper panel of **Fig. 5f**), which is simplified to a zig-zag shape. It is then clear that an overshoot/undershoot can only happen at an ionic bottleneck in SOEC. Moreover, at any electronic bottleneck that does not coincide with an ionic bottleneck, by applying $\delta\mu_{O_2} = -4\delta\mu_e$ we can see that the $\mu_{O_2}$ jump is rightward in both SOFC and SOEC (lower panels of **Fig. 5e-f**), so no overshoot/undershoot should ever arise there.

Next, we recall the analogy between SOFC/SOEC and rechargeable batteries—by mapping $O_2$ to $R$, $O^{2-}$ to $R^+$, and electron to hole. This means that the schematics in **Fig. 5** can also be applied to Li/Na batteries, which is especially valuable because no $\sigma_e$ and $\sigma_h$ data are currently available in any solid electrolyte of rechargeable Li/Na batteries to perform similar numerical calculations. The analogy allows us to immediately conclude that $\mu_R$ overshoot/undershoot can only arise from an ionic bottleneck in the charge mode but not in the discharge mode. In the case when the overshoot exceeds the boundary potential of the negative (Li/Na metal) electrode, metal precipitation is expected on the internal grain boundaries of the solid electrolyte as illustrated in the schematic for a Na-S battery in **Fig. 2b**.

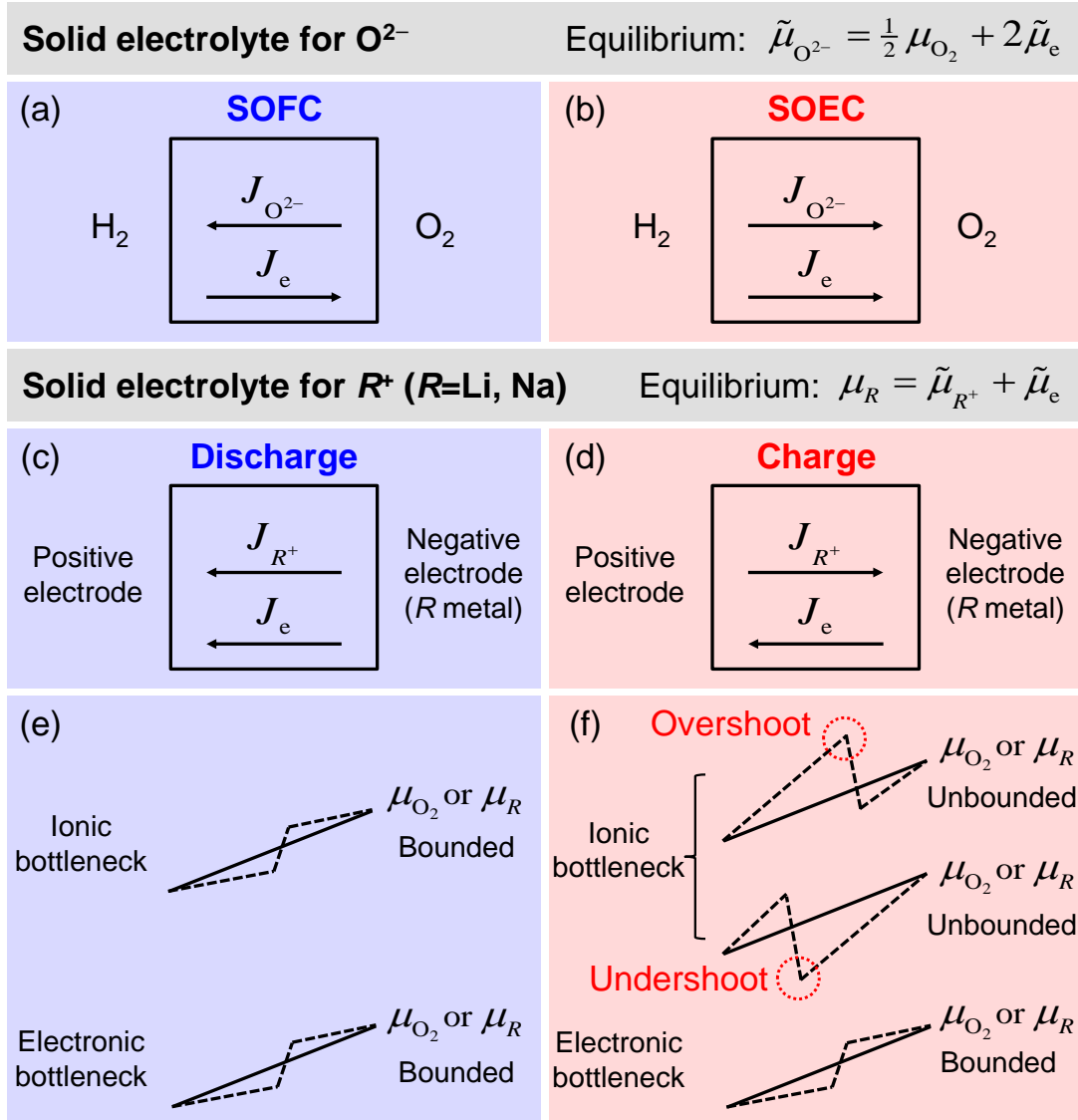

**Figure 5 Potential overshoot/undershoot in solid electrolyte.** Schematics of solid electrolyte in (a) SOFC and (b) SOEC showing directions of ionic flow $J_{O^{2-}}$ and electron flow $J_e$ (hole flow opposite to $J_e$). Analogous schematics of solid electrolyte for LIB and Na-S battery in (c) discharge and (d) charge. (e-f) Straight and zig-zag $\mu_{O_2}$ distributions representing simpled $\mu_{O_2}$ profiles in **Fig. 3a-b** and **4a-b**. Non-linear base-line $\mu_{O_2}$ profiles shown as straight (solid) lines between boundary potentials. Non-linear $\mu_{O_2}$ jumps across bottleneck shown as zig-zag (broken) lines in (upper panel) ionic channel and (lower panel) electronic channel.

Overshoot/undershoot encircled in red in (f) corresponds to similar overshoot/undershoot encircled in purple in **Fig. 3a/b**. Schematics in (e-f) also hold for $\mu_R$ in battery. Notes: (i) total current flow is completely dominated by ion flow in these good electrolytes, but for clarity the magnitudes of $J_e$ arrows have been grossly exaggerated; (ii) devices have no leak, hence no flow of respective underlying molecular/atomic species ($O_2$ and $R$) in these solid electrolytes.

**IV. Tuning electronic conductivity to mitigate potential overshoot/undershoot**

Since we attributed the largest $\mu_{O_2}$ jump in **Fig. 1b-c** to the sharp $(\sigma_e + \sigma_h)_{min}$ at the electronic bottleneck, it stands to reason that if the sharp minimum is smeared, then the largest $\mu_{O_2}$ jump and the consequent $\mu_{O_2}$ overshoot/undershoot at ionic bottlenecks can be avoided. To test this idea, we repeated the calculations in **Fig. 3** for another electrolyte in SOEC that has the same $\sigma_{O^{2-}}$ as YSZ but with a smoother $(\sigma_e + \sigma_h)_{min}$ (**Fig. 6a**). This may be achieved by doping YSZ with a donor or acceptor that has a redox-insensitive (extrinsic) conductivity. As shown in **Fig. 6b**, even though doping still keeps $\sigma_e + \sigma_h$ miniscule, smaller than $\sigma_{O^{2-}}$ by more than 1,000 times at all $\mu_{O_2}$, it still smooths out and even "linearizes" the $\mu_{O_2}$ profiles, thus relieves most of the electrolyte from the influence of extreme oxidizing/reducing conditions at the electrolyte/electrode interfaces. As most part of the electrolyte no longer bears the burden of the electrode overpotentials, it should no longer experience any $\mu_{O_2}$ overshoot/undershoot beyond the boundary potentials. This will remove the possibility of forming unexpected phases inside the electrolyte. Although the calculations were performed for a doped YSZ, in view of the analogy of Eq. (1-4) and **Fig. 5**, we expect similarly minor donor/acceptor doping will also benefit solid electrolytes in rechargeable Li/Na batteries.

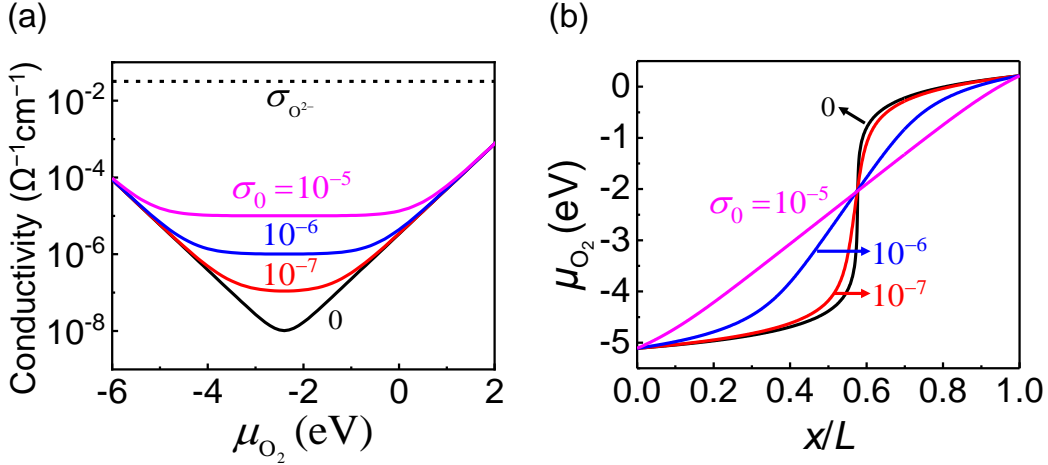

**Figure 6 Smoothed $(\sigma_e + \sigma_h)_{min}$ linearizes $\mu_{O_2}$ distribution inside solid electrolyte.** (a) Total electronic conductivity including intrinsic $\mu_{O_2}$-dependent $\sigma_e + \sigma_h$ (black solid curve, same as in **Fig. 1a**) and extrinsic $\mu_{O_2}$-independent $\sigma_0$ =$10^{-7}$, $10^{-6}$, or $10^{-5}$ $\Omega^{-1}$cm$^{-1}$. (b) Calculated $\mu_{O_2}$ distributions in single-crystal YSZ SOEC under various $\sigma_e + \sigma_h + \sigma_0$. Conditions: 800°C and −1 A/cm$^2$ with $\mu'_{O_2} = -5.11$ eV ($PO_2$=10$^{-24}$ atm) at $x$=0, $\mu''_{O_2} = 0.21$ eV ($PO_2$=10 atm) at $x$=L, L=10 μm. As $\mu_{O_2}$ jump progressively disappears, ionic current continues to dominate as indicated by calculated ionic vs. electronic contribution to total current density: 0.9986 vs. 0.0014 for $\sigma_0 = 0$, 0.9985 vs. 0.0015 for $\sigma_0 = 10^{-7}$, 0.9973 vs. 0.0027 for $\sigma_0 = 10^{-6}$, and 0.9851 vs. 0.0149 for $\sigma_0 = 10^{-5}$.

## V. Model predictions

Our results lead to highly specific predictions as listed below. They may be used as the "selection rules" to select our potential-overshoot/undershoot mechanism over other possible mechanisms as the mechanism responsible for the formation of an unexpected neutral phase in an electrochemical device.

(a) Mode: It occurs in SOEC but not in SOFC, and in charging but not in discharging.

(b) Location: It occurs inside the electrolyte, but not at the electrolyte/electrode interface. (See **Appendix 7** of Ref. [25] for comparison with work of Yoo *et al*. [40, 41])

(c) Orientation: It occurs on grain boundaries perpendicular to the ionic current, but not those parallel to the current. This is because current can bypass the latter, whereas bypassing the former will cause current concentration nearby, which aggravates phase formation, even "chain reactions", on other transverse boundaries.

(d) Polarity: Oxygen bubbles [4, 7-10] form on the oxygen electrode side but not the hydrogen electrode side, reduction voids [34, 35] form on the hydrogen electrode side but not the oxygen electrode side, and metal islands [16-19] form on the negative electrode side but not the positive electrode side.

(e) Spatial extension: Prevalent phase formation is accompanied by a redox partition demarcating a sharp (and often extreme) $\mu_{O_2}$ or $\mu_R$ transition, with the severely reduced side possibly having a uniform metal precipitation [17] or grain growth [33, 34, 42] (due to reduction-enhanced cation diffusion [43]). The severely oxidized side may suffer from other instability such as phase dissolution.

(f) Time: "Exceptions" to the above rules may occur when unexpected phase formation has already reached an advanced stage to cause widespread damage, e.g., the electrolyte/electrode interface may delaminate by linking O$_2$ bubbles, and dendrites may form by linking metal islands.

(g) Mitigation: Phase formation can be avoided by minute doping of a donor/acceptor that has a redox-insensitive conductivity, or dopant that enhances grain boundary diffusivities (e.g., by lessening the space charge [26, 27]). Since an incubation time is needed for nucleation, alternating the mode of operation [4] or switching between high and low operational current (charge rate) to eliminate the overshoot/undershoot before the unexpected phase is nucleated should be effective in mitigating damage. Smaller grains may help divert the flow around low-diffusivity boundaries, and reducing diffusion anisotropy, e.g., using a glassy electrolyte, can obviate ionic bottlenecks.

## VI. Comparison with experimental observations

Below we discuss the observations of damages in YSZ (SOEC) and BASE/LLZO electrolytes summarized in **Table 1**, *vis a vis* the predictions in **Section V**. (NASICON develops similar damage as BASE.)

On YSZ SOEC, there are the long-standing findings of oxygen bubbles on YSZ grain boundaries near the oxygen electrode, especially at high current densities and during long operations. [4, 7-10] They were attributed to the electrode overpotential at the oxygen electrode, which elevates the oxygen potential in the nearby YSZ electrolyte. [4, 7, 8, 22-24] However, such a mechanism would predict bubble formation at all grain boundaries and especially at their triple/quadruple-grain junctions, because with a uniform driving force these junctions are the most favorable nucleation sites. Yet most bubbles and their coalescence cracks were found on transverse boundaries, displaying a propensity for longer boundaries and no preference for triple/quadruple junctions. Near the hydrogen electrode, oxygen-vacancy cavities (and their coalescence cracks) also preferentially form along transverse grain boundaries. [34, 35] Unlike SOEC, SOFC does not suffer from the above damages. Indeed, even though most time is spent in the SOEC mode, periodic reversal of the current direction to briefly operate in the SOFC mode can suppress oxygen bubble formation. [4] As bubbles accumulate, the area of intact grain boundaries decreases, so it is forced to support a higher local current density to sustain the same $O^{2-}$ flow, which in turn necessitates a larger $\mu_{O_2}$ jump and an increased overall resistance. [4] This phenomenon is illustrated in the sample calculations in **Fig. 7**. Lastly, at higher testing temperatures, a sharp boundary with grains of markedly different (10 times) sizes on two sides was found to delineate the redox boundary [42], which coincides with $(\sigma_e + \sigma_h)_{min}$ and the corresponding $\mu_{O_2}$ transition. The above observations are all consistent with our predictions (a-g). In the on-line Ref. [25] we show that, without affecting the apparent device resistance, the minuscule electronic conductivity in YSZ can supply the necessary electronic current across the nominally ionic solid electrolyte to charge-compensate the formation of neutral phases from charged species ($O^{2-}$ ion or its lattice defect $V_O^{\bullet\bullet}$).

**Table 1 Summary of experimental observations**

| Materials | Properties | Function | Mode | Experimental observations |
|---|---|---|---|---|
| YSZ | Fast $O^{2-}$ conductor, electronic insulator | Solid electrolyte | SOEC | Oxygen bubbles and cracks on oxygen-electrode side, preferentially at transverse grain boundaries [4, 7-10]; oxygen electrode delamination; reduction cavities and cracks on hydrogen-electrode side, preferentially at transverse grain boundaries [34, 35]; sharp boundary between reduced and unreduced regions [33, 42]. |
| Beta alumina | Fast $Na^+$ conductor, electronic insulator | Solid electrolyte | Charging | Na metal precipitation and micro-cracks at transverse grain boundaries, voids and cracks on Na-metal-electrode side; uniform, precipitation layer propagating from Na-side; sharp boundary between reduced and unreduced regions; graphite (current connector for sulfur electrode) imprint. [17-19] |
| LLZO | Fast $Li^+$ conductor, electronic insulator | Solid electrolyte | After cycling | Li metal precipitation at grain boundaries away from Li-metal electrode and without apparent Li dendrite [14-16]; uniform Li-metal precipitation with constant Li concentration near anode [16]. |
| SEI on graphite negative electrode | Fast $Li^+$ conductor, electronic insulator | SEI | After cycling | Activity of metallic Li peaking at 20-30 nm beneath the electrolyte-SEI surface, exceeding boundary values at the surface and in bulk graphite. [50] |
| Porous graphene network | Mixed $Li^+$ and electron conductor | Negative electrode | Charging | Li metal precipitation when tested between 3 V and 0.03 V vs. Li/$Li^+$ in half-cell configuration using Li metal reference electrode. [51] |
| $LiNi_{0.80}Co_{0.15}Al_{0.05}O_2$, $Li_{1.2}Ni_{0.13}Co_{0.13}Mn_{0.54}O_2$ | Mixed $Li^+$ and electron conductor | Positive electrode | Charging | $O_2$, $CO_2$ gas evolution. [52, 53] |

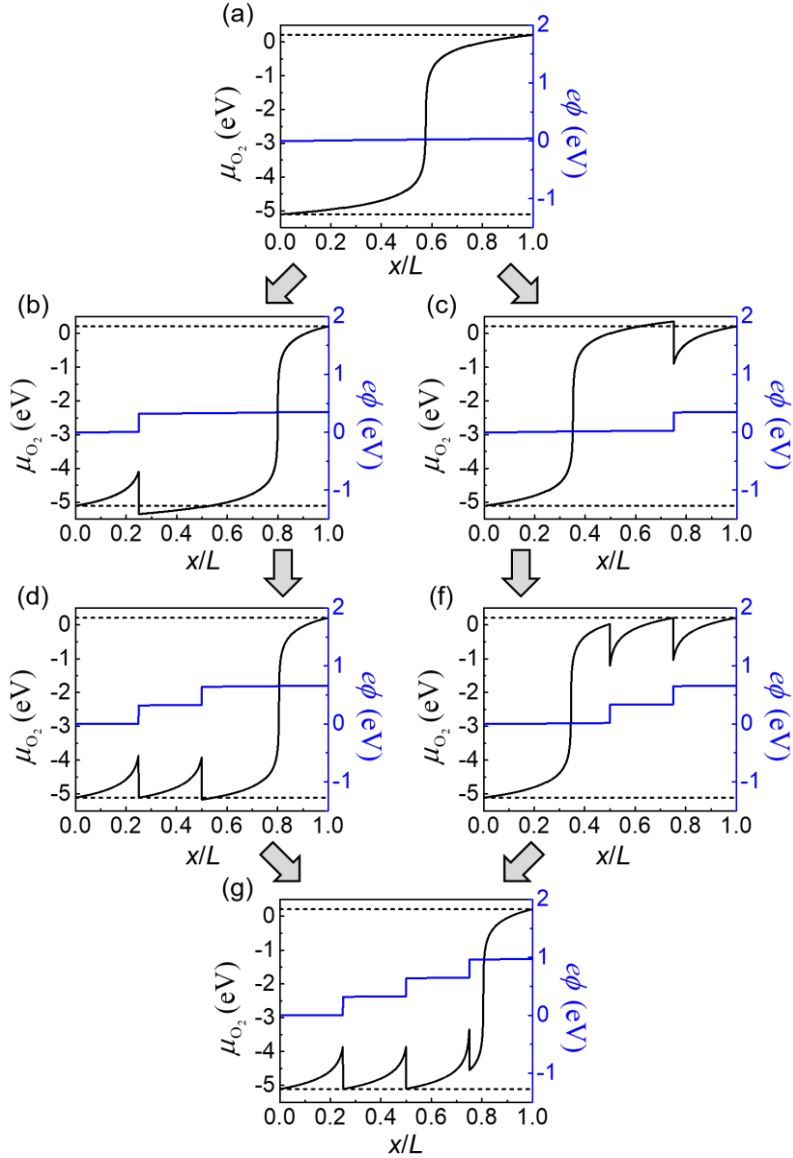

**Figure 7 Calculated $\mu_{O_2}$ and $e\phi$ inside "polycrystalline" solid electrolytes in SOEC.** Polycrystalline YSZ electrolyte in SOEC at −1 A/cm² with grain boundaries at ¼L, ½L and ¾L from left and relative $\sigma_{O^{2-}}$ of (a) all boundaries set at $\sigma_{O^{2-}}^{GB}/\sigma_{O^{2-}}^{L}$ =10⁻², (b) first boundary set as 10⁻⁴, (c) last boundary set as 10⁻⁴, (d) first and second boundaries set as 10⁻⁴, (e) second and last boundaries set as 10⁻⁴, and (f) all boundaries set as 10⁻⁴. Other conditions: 800°C with $\mu'_{O_2}$ = −5.11 eV ($PO_2$=10⁻²⁴ atm) at x=0, $\mu''_{O_2}$ = 0.21 eV ($PO_2$=10 atm) at x=L, L=10 μm, t=10 nm. Note: total voltage drop from x=0 to x=L increases with the number of very-low-$\sigma_{O^{2-}}$ boundaries.

On rechargeable Na-S batteries, there are the long-standing observations of failure-causing internal Na metal deposits in BASE during Na-charging. [17-19] Often found at cracks and voids thus motivated the initial explanation based on electrical-field/current concentrations, they also formed at triple-grain junctions and two-grain boundaries—primarily ones transverse to the Na$^+$ flow. Since Na$^+$ in BASE moves inside the open "galleries" between spinel blocks [44], Na$^+$ conduction is highly anisotropic and much slower in some grains and across most grain boundaries [45, 46]. Therefore, ionic bottlenecks can explain the deposits at cracks, voids and transverse grain boundaries. Interestingly, next to the (sulfur) positive electrode, BASE dissolution causing graphite (serving as a charge connector for the sulfur electrode) imprint was also observed. [44] This implicates a $\mu_{Na}$ undershoot so severe that BASE becomes unstable, thus undergoing dissolution. In rechargeable LIB, rather similar observations of failure-causing internal Li metal deposits were recently reported for LLZO [14-16, 47], especially along transverse grain boundaries on the negative electrode side. Finally, evidence of a sharp redox boundary was seen in BASE with the reduced side next to the negative electrode decorated by a relatively uniform precipitation of Na metal without preference to the transverse grain boundaries [44]; same observations were also recently made in LLZO [16]. With time, the electrolyte may be shorted by the growing metal precipitates, which moves the effective negative electrode and the redox front inward. These observations are all consistent with our predictions (a-f); mitigating them by employing (g) has not been reported yet.

We believe these predictions can also explain observations of metal deposits in solid electrolyte interphase (SEI) that forms between the negative electrode and the liquid electrolyte, and instability of cathode electrolyte interphase (CEI) that forms between the positive electrode and the liquid electrolyte in LIBs. [48, 49] In terms of transport properties, SEI and CEI are relatively good ion conductors and nearly electron insulators, so they may be regarded as good electrolytes. Despite their very thin thickness, they must support extreme redox potentials and are chemically and microstructurally heterogeneous. So they likely contain some ionic/electronic bottlenecks with $\mu_R$ jumps exceeding the boundary potentials. **Appendix 3** and **Fig. A1** provide a more detailed analysis, predictions and discussion of experimental observations of metal deposits in SEI and CEI in LIB [48, 50] (also summarized in **Table 1.**)

## VII. Concluding remarks and implications beyond solid electrolytes

Our thermodynamic analysis has discovered that transport bottlenecks in electronic/ionic channels can transfer and amplify the extreme overpotentials of the electrodes deep inside the solid electrolyte, which is the root cause of unexpected phase formation and accelerated degradation. The theory is consistent with the common knowledge that high current densities are more damaging and low temperature operations (such as running a battery in a frigid weather) makes matter worse, but it also provide specific predictions that help identify the conditions conducive for such damage and means to mitigate it. For example, glass and liquid electrolytes that have no structural heterogeneities above the atomic scale should have no ionic bottlenecks, so they are better choices, although they will still suffer from the same electronic bottleneck as shown in **Fig. 1a**. The analysis can be extended beyond "good" electrolytes to LIB electrodes and other mixed ionic and electronic conductors as outlined in **Appendix 4** and **Fig. A2**, which again predicts $\mu_{Li}$ overshoot/undershoot under specific conditions consistent with the recent observations of unexpected Li metal precipitation and $O_2/CO_2$ evolution in LIB electrodes (also summarized in **Table 1**). [51-53] These findings have exemplified the power of thermodynamics analysis under the assumptions of steady state and local equilibrium. [54] Future extensions possibly aided by first-principles calculations [55] and non-steady-state/non-equilibrium analysis together with model experiments will further our understanding and help improve the reliability of electrochemical energy devices.

**Acknowledgements**

This work was supported by the Department of Energy (BES grant no. DEFG02-11ER46814) and used the facilities (LRSM) supported by the U.S. National Science Foundation (grant no. DMR-1120901).

## Appendix 1. Solving $\mu_{O_2}$ and other potentials in YSZ electrolyte

With $\mu'_{O_2} < \mu''_{O_2}$ at the two boundaries $x=0$ and $x=L$ (where $L$ is the thickness of the electrolyte) in **Fig. 3**, we will solve the steady-state distributions of potentials in the YSZ slab under a prescribed total electrical current density $j_{\text{total}}$, defined as positive if it flows from left to right. We follow the method and notation described in Ref. [33]. Excluding internal reactions, we consider four species: oxygen ion $O^{2-}$, oxygen molecule $O_2$, electron e and hole h. Under local equilibrium, the two chemical reactions

$$O^{2-} = \tfrac{1}{2} O_2 + 2e \qquad (A1)$$

$$e + h = \text{nil} \qquad (A2)$$

relate the three electrochemical potentials and the chemical potential $\mu_{O_2}$ of $O_2$ by

$$\tilde{\mu}_{O^{2-}} = \tfrac{1}{2}\mu_{O_2} + 2\tilde{\mu}_e \qquad (A3)$$

$$\tilde{\mu}_e + \tilde{\mu}_h = 0 \qquad (A4)$$

Because no mobile $O_2$ molecule is allowed in a good device, the only fluxes are those of charged species

$$j_{O^{2-}} = \frac{\sigma_{O^{2-}}}{2e} \frac{d\tilde{\mu}_{O^{2-}}}{dx} \qquad (A5)$$

$$j_e = \frac{\sigma_e}{e} \frac{d\tilde{\mu}_e}{dx} \qquad (A6)$$

$$j_h = -\frac{\sigma_h}{e} \frac{d\tilde{\mu}_h}{dx} = \frac{\sigma_h}{e} \frac{d\tilde{\mu}_e}{dx} \qquad (A7)$$

where $\sigma_i$ denotes the conductivity of species $i$, which varies as a function of local $\mu_{O_2}$ and structure, i.e., it matters whether it is in the lattice or at a grain boundary. Since electrons and holes can be generated and annihilated via Eq. (A2) and they follow Eq.

(A4, A6-7), their current densities combine to a total electronic current density $j_{eh}$

$$j_{eh} = j_e + j_h = \frac{\sigma_e + \sigma_h}{e} \frac{d\tilde{\mu}_e}{dx} \quad (A8)$$

At the steady state and without crosstalk between ionic and electronic currents, the ionic current density $j_{O^{2-}}$ and the electronic current density $j_{eh}$ must each remain constant throughout the electrolyte. Therefore, derivatives of $\tilde{\mu}_i$ can be expressed in terms of $j_{O^{2-}}$ and $j_{eh}$ via Eq. (A5) and (A8),

$$\frac{d\tilde{\mu}_{O^{2-}}}{dx} = 2e \frac{j_{O^{2-}}}{\sigma_{O^{2-}}} \quad (A9)$$

$$\frac{d\tilde{\mu}_e}{dx} = -\frac{d\tilde{\mu}_h}{dx} = e \frac{j_{eh}}{\sigma_e + \sigma_h} \quad (A10)$$

Using Eq. (A9-A10) and Eq. (A3), we obtain, after rearrangement,

$$f(x/L) \equiv \frac{d\mu_{O_2}}{d(x/L)} = 4eL\left(\frac{j_{O^{2-}}}{\sigma_{O^{2-}}} - \frac{j_{eh}}{\sigma_e + \sigma_h}\right) \quad (A11)$$

Therefore, a steep change in $\mu_{O_2}$ can arise from either a small $\sigma_{O^{2-}}$ (in the ionic channel) or a small $\sigma_e + \sigma_h$ (in the electronic channel) regardless whether $j_{eh} \ll j_{O^{2-}}$ or not.

For convenience, we now write the ratio of ionic to total current density $j_{total}$ as $t_i = j_{O^{2-}}/j_{total}$. Here and below, $t_i$ should not be confused with grain boundary thickness, which is $t$. In the SOFC mode, $j_{O^{2-}}$ and $j_{eh}$ are in opposite directions and $j_{total} = j_{O^{2-}} + j_{eh} = j_{O^{2-}} - |j_{eh}|$. Therefore, Eq. (A11) can be written as

$$f(x/L) = 4eLj_{total}\left(\frac{t_i}{\sigma_{O^{2-}}} + \frac{t_i - 1}{\sigma_e + \sigma_h}\right) \quad (A12)$$

where $t_i - 1$ is the ratio of electronic current density (carried by electrons and holes) to $j_{total}$. In the SOEC mode, $j_{O^{2-}}$ and $j_{eh}$ are along the same directions and

$j_{\text{total}} = j_{O^{2-}} + j_{eh}$. Therefore, Eq. (A11) can be written as

$$f(x/L) = 4eL|j_{\text{total}}|\left(\frac{1-t_i}{\sigma_e + \sigma_h} - \frac{t_i}{\sigma_{O^{2-}}}\right) \quad (A13)$$

where $1-t_i$ is the ratio of electronic current density to $j_{\text{total}}$. For completeness, we also give the corresponding form in the open-circuit-voltage (OCV) mode, which is

$$f(x/L) = 4eLj_0\left(\frac{1}{\sigma_{O^{2-}}} + \frac{1}{\sigma_e + \sigma_h}\right) \quad (A14)$$

where $j_0 = j_{O^{2-}} = -j_{eh}$ denotes the absolute value of the ionic and electronic current density at OCV. Therefore, in all three forms, Eq. (A12-A14), there is only one unknown constant to solve, which is $t_i$ for Eq. (A12-A13) and $j_0$ for Eq. (A14). This is done by satisfying the boundary condition

$$\int_{\mu'_{O_2}}^{\mu''_{O_2}} \frac{d\mu_{O_2}}{f(x/L)} = 1 \quad (A15)$$

Finally, the oxygen potential distribution can be obtained by integrating $f(x/L)$ from $x=0$ to arbitrary $x$

$$x = L\int_{\mu'_{O_2}}^{\mu_{O_2}} \frac{d\mu_{O_2}}{f(x/L)} \quad (A16)$$

Solutions of all three modes have already been described in Ref. [33]. Importantly, we note that with the same $\mu'_{O_2} < \mu''_{O_2}$ the $\mu_{O_2}$ profiles are all similar. This confirms our prediction that the profile is primarily determined by the electronic current, which has the same direction in all three modes. (See **Fig. 5** and **Appendix 3** for directions of $j_{eh}$. Under OCV, $j_{eh} = -j_{O^{2-}}$.)

To allow for special conductivity at special locations, such as grain boundaries, we use a set of discrete coordinates $x_n$ where the oxygen potential is $\mu_n$. Instead of Eq.

(A15), the boundary condition can now be expressed as

$$\mu''_{O_2} - \mu'_{O_2} = \sum_{n=0}^{N-1} f(x_n/L)\left(\frac{x_{n+1}-x_n}{L}\right) \qquad (A17)$$

In the above, $n$ runs from 0 to $N$ with 0 corresponding to $x=0$ and $N$ to $x=L$ in the continuum description. Likewise, instead of Eq. (A16), the oxygen potential distribution can be obtained from

$$\mu_{O_2} = \mu'_{O_2} + \sum_{n=0}^{m-1} f(x_n/L)\left(\frac{x_{n+1}-x_n}{L}\right) \qquad (A18)$$

After obtaining $\mu_{O_2}(x)$, $\tilde{\mu}_{O^{2-}}(x)$ and $\tilde{\mu}_e(x)$ can be calculated within an integration constant by integrating Eq. (A9-10). (The two integration constants are related to each other by Eq. (A3)). This remaining constant corresponds to an arbitrary reference potential of $\phi$, which is set by letting $\tilde{\mu}_{O^{2-}}(x=0)=0$. Lastly, because $\mu_{O^{2-}}$ is a constant in YSZ, $-2e\phi$ differs from $\tilde{\mu}_{O^{2-}}$ by an additive constant only, which is again fixed by letting $\tilde{\mu}_{O^{2-}}(x=0)=0$ in this work. The solution is now complete once $\sigma_i$ is known at every $x_n$. Such information comes from either the local $\mu_{O_2}$ or our knowledge of the microstructure (e.g., the locations of low-$\sigma_{O^{2-}}$ grain boundaries). Specifically, (i) oxygen conductivity is independent of $\mu_{O_2}$ (~0.03 $\Omega^{-1}\text{cm}^{-1}$ at 800 °C in the lattice), and the $\mu_{O_2}$ dependence of electron and hole conductivity is that of Park and Blumenthal [28] (see conductivity data in **Fig. 1**); and (ii) the ratio of $\sigma_{O^{2-}}^{GB}/\sigma_{O^{2-}}^{L}$ is known from AC impedance spectroscopy (measured at 250-500°C and extrapolated to 800°C provided $\sigma_{O^{2-}}^{GB} = \sigma_{O^{2-}}^{L}/160$ according to Guo and Maier [26, 27]). The ratio could be much lower for some grain boundaries because AC impedance spectroscopy largely overlooks the less conductive grain boundaries.

**Appendix 2. Directions of ion and electron flows**

Below, Δ refers to the difference of values between the two electrode-electrolyte interfaces. Consider the YSZ electrolyte in **Fig. 5a-b** and write the potential differences between the oxygen electrode and the hydrogen electrode as $V_{th} = \frac{\Delta \mu_{O_2}}{4e} > 0$, $V = -\frac{\Delta \tilde{\mu}_e}{e}$ and $\Delta \tilde{\mu}_{O^{2-}}$, following the notation of Riess [56]. They satisfy the equilibrium condition $\Delta \tilde{\mu}_{O^{2-}} = \frac{1}{2} \Delta \mu_{O_2} + 2 \Delta \tilde{\mu}_e = 2e(V_{th} - V)$. So the free variable $V$ would determine the modes of operation and directions of flows: the direction of $O^{2-}$ flow by $\Delta \tilde{\mu}_{O^{2-}} = 2e(V_{th} - V)$, and that of electron flow by $\Delta \tilde{\mu}_e = -eV$.

When $V > V_{th} > 0$, $O^{2-}$ flows from the hydrogen electrode to the oxygen electrode, which is the case of SOEC. Conversely, when $V < V_{th}$, $O^{2-}$ flows from the oxygen electrode to the hydrogen electrode, which is the case of SOFC. For electrons, throughout the SOEC range, $V > 0$, so electrons flow from the hydrogen electrode to the oxygen electrode; in the SOFC range up to $0 < V < V_{th}$, electrons also flow from the hydrogen electrode to the oxygen electrode. When $V = 0$ is reached, it corresponds to short-circuiting the external circuit, thus the range beyond which is of no practical interest. This explains all the directions of $O^{2-}$ ion and electron flows in **Fig. 5a-b**.

The same argument applies to solid electrolytes for Li$^+$ and Na$^+$, with potential differences between two electrodes $V_{th} = \frac{\Delta \mu_{Li/Na}}{e} > 0$, $V = -\frac{\Delta \tilde{\mu}_e}{e}$, and $\Delta \tilde{\mu}_{Li^+/Na^+} = \Delta \mu_{Li/Na} - \Delta \tilde{\mu}_e = e(V_{th} + V)$. During charging, $-V > V_{th} > 0$, so ion flows to the right (i.e., Li/Na metal electrode) and electrons to the left (i.e., cathode). During discharging, $0 < -V < V_{th}$, so both ions and electrons flow to the left (i.e., cathode) from the Li/Na metal electrode. When $V = 0$ is reached, it corresponds to short-circuiting the external circuit thus the range beyond which is of no practical interest. This explains all the directions of ion and electron flows in **Fig. 5c-d** for Li$^+$ and Na$^+$ electrolytes.

**Appendix 3. Damage development in SEI and CEI in batteries**

LIBs and other advanced batteries often operate at extreme potentials beyond the thermodynamic stability window of liquid organic electrolytes. As a result, a passivation layer called SEI/CEI typically forms at the electrode surfaces. As proposed by Peled [48] and generalized by others [49], SEI and CEI are solid products from side reactions between electrodes and electrolytes, and a good SEI/CEI should conduct Li$^+$ but not electrons. Applying **Fig. 5** (with more details shown in **Fig. A1**) to SEI/CEI, we expect $\mu_R$ overshoots/undershoots at ionic bottlenecks during charging. In support of this, on the more reducing side (i.e., SEI), internal Li metal deposits at 20-30 nm away from the surface were found in the SEI on the graphite anode after cycling, indicating the internal $\mu_{Li}$ there had reached the level of Li metal and exceeded SEI's upper boundary $\mu_{Li}$ (i.e., $\mu_{Li}$ of lithiated graphite). [50] This compromises the integrity of SEI as a membrane of selective transport. (Paled speculated that internal Li metal precipitation arising from excessive local heating may cause unstable SEI. [48]) The situation at CEI could be worse: it is on the more oxidizing side, so highly destructive internal $O_2/CO_2$ bubbles could form at any severe undershoot of internal $\mu_{Li}$.

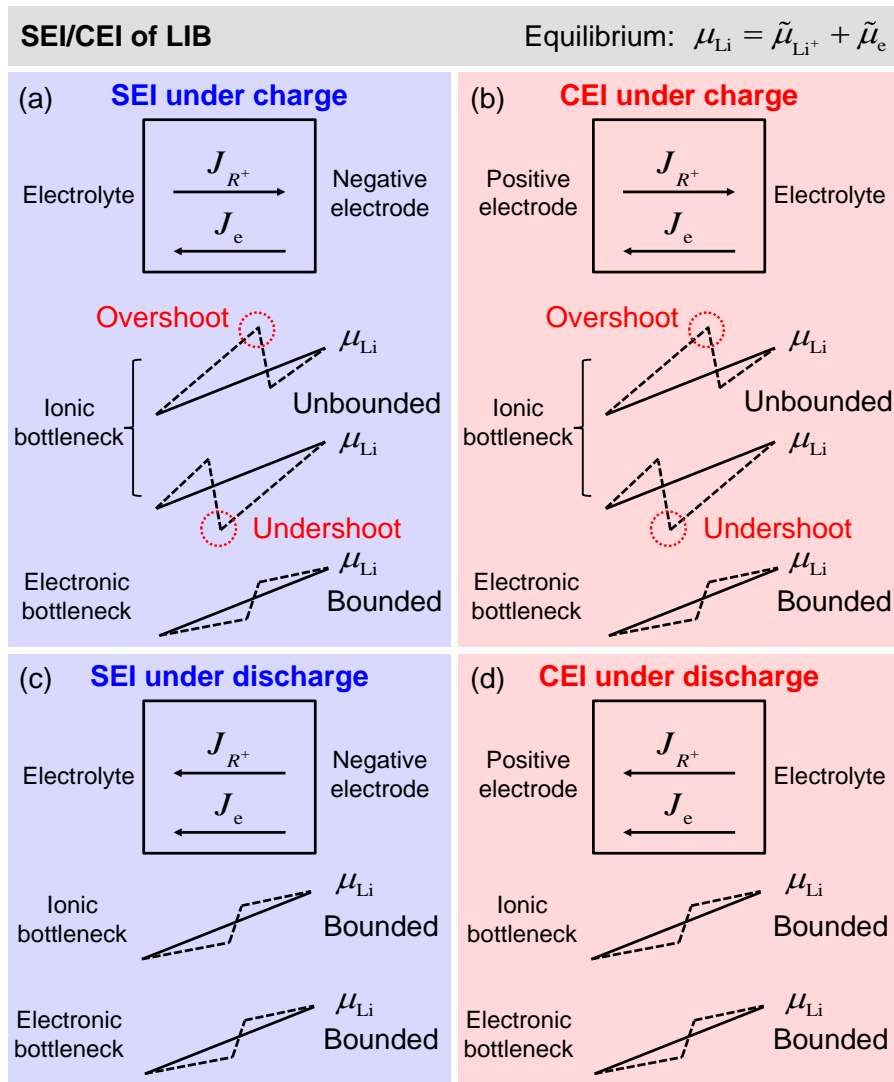

**Figure A1 Potential overshoot/undershoot in SEI/CEI.** Schematics of charge and discharge of (a & c) solid electrolyte interface (SEI) between negative electrode and electrolyte and (b & d) cathode electrolyte interface (CEI) between positive electrode and electrolyte, showing directions of ionic flow $J_{Li^+}$ and electron flow $J_e$ (hole

flow in the opposite direction). Also shown are schematic profiles of $\mu_{Li}$ jump across a bottleneck in ionic channel (upper panel) and electronic channel (lower panel). As in **Fig. 5**, nonlinear base-line $\mu_{Li}$ are simplified to straight lines between boundary potentials, and $\mu_{Li}$ jumps at bottlenecks simplified to zig-zags. If these zig-zags exceed boundary values, their peaks are circled and termed as overshoot/undershoot. The lengths of $J_{Li^+}$ and (grossly exaggerated) $J_e$ arrows are not accurate and meant for schematics only.

**Appendix 4. Damage development in LIB Electrodes**

Unexpected Li metal precipitation has been observed in LIB anodes when they are cycled above 0 V vs. Li/Li$^+$, suggesting a $\mu_{Li}$ overshoot. For example, Li metal precipitation occurred inside a porous graphene network on the negative electrode side, tested against a Li metal electrode in the half-cell configuration between 3.0 V and 0.03 V vs. Li/Li$^+$. [51] Conversely, the frequent report of $O_2/CO_2$ evolution from the positive electrodes [52, 53] may be taken as tentative evidence of $\mu_{Li}$ undershoots on the more oxidizing side. Good redox-active advanced electrodes should have facile electronic conductivity, and many of them also enjoy a broad range of stoichiometry and valence states for transition-metal cations or even anions. Therefore, an extremely simplified first approximation for the electrode is to assume a constant $\tilde{\mu}_e$ instead of a constant $\mu_{Li^+}$ as before. This implies any gradient in the electrochemical potential of Li$^+$, $\tilde{\mu}_{Li^+} = \mu_{Li^+} + e\phi$, will be reflected in a gradient of $\mu_{Li}$. Since Li$^+$ diffusion in the electrode is likely to be slower than that in the electrolyte, one may expect a graded $\tilde{\mu}_{Li^+}$ to develop in the electrode, which suffices to specify its $\mu_{Li}$ as shown in **Fig. A2** (i.e., the Li$^+$ rate-limiting case). This knowledge allows us to predict their $\mu_{Li}$ overshoot/undershoot at an electronic bottleneck. On the other hand, some redox-active electrodes in use, e.g., LiFePO$_4$, do suffer from relatively poor electronic conductivity and need a conductive coating, so the other extreme approximation is to assume a constant $\tilde{\mu}_{Li^+}$. This implies any gradient in $\tilde{\mu}_e$ will be reflected in a gradient of $\mu_{Li}$, and as shown in **Fig. A2**, their overshoot/undershoot now occurs at an ionic bottleneck.

Further considering the highly oxidizing condition (low $\mu_{Li}$) on the positive electrode side especially at high voltages and *vice versa* on the negative electrode side at low voltages, we believe the most significant scenarios are a $\mu_{Li}$ undershoot in the former, and a $\mu_{Li}$ overshoot in the latter. For the positive electrode, a $\mu_{Li}$ undershoot and O$_2$/CO$_2$ formation is likely to happen (i) at an ionic bottleneck in the electron rate-limiting case, or (ii) at an electronic bottleneck in the Li$^+$ rate-limiting case during delithiation (i.e. charging) at high voltages. For the negative electrode, a $\mu_{Li}$ overshoot and Li metal formation is likely (iii) at an ionic bottleneck in the electron rate-limiting case, or (iv) at an electronic bottleneck in the Li$^+$ rate-limiting case during lithiation (i.e. charging) at low voltages. Therefore, our theory (iii or iv) can explain why Li metal precipitates in Ref. [51]. To definitively correlate O$_2$/CO$_2$ evolution [52, 53] to our theory (i or ii), we recommend the following experiment: in a half-cell configuration and with the same cut-off voltage, observe faster O$_2$/CO$_2$ gas generation starting at a lower voltage when the cell delithiates at a faster charge rate.

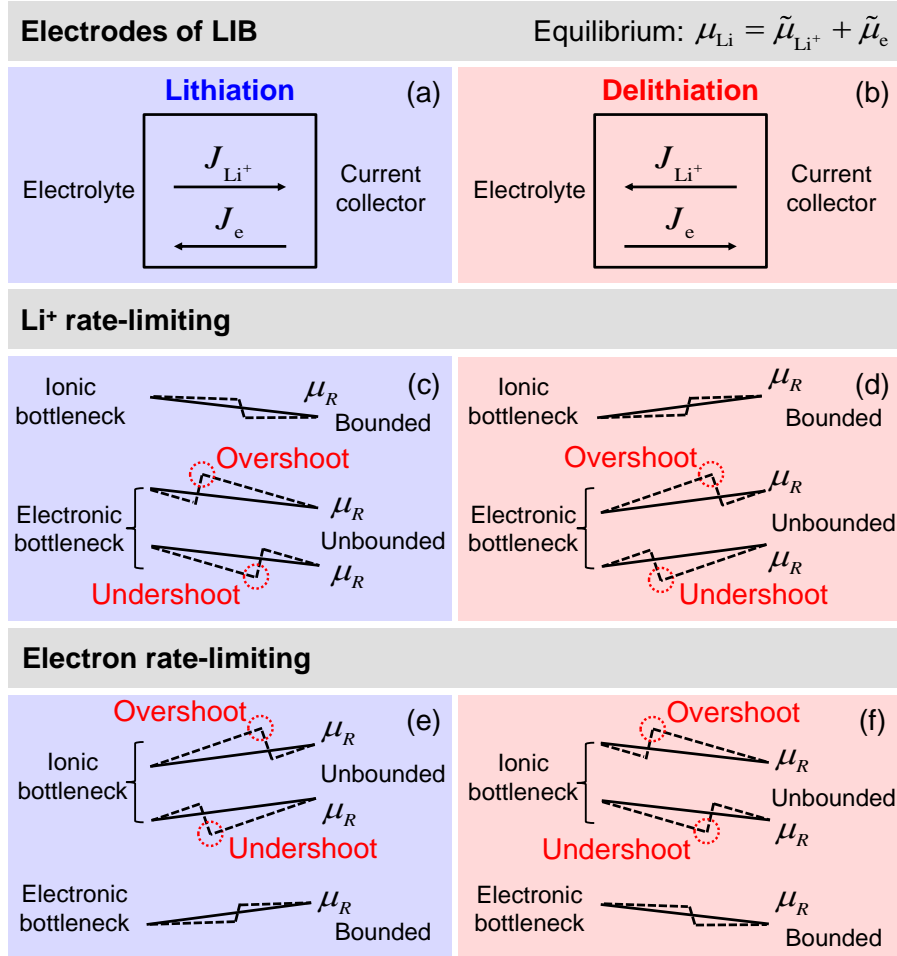

**Figure A2 Potential overshoot/undershoot in active electrodes.** Schematics of (a) lithiation and (b) delithiation of electrodes for LIB showing directions of ionic flow $J_{Li^+}$ and electron flow $J_e$ (hole flow in the opposite direction). Two extreme cases considered: (c-d) Li$^+$ rate-limiting with constant $\tilde{\mu}_e$, and (e-f) electron rate-limiting with constant $\tilde{\mu}_{Li^+}$. Also shown are schematic profiles of $\mu_{Li}$ jump across a bottleneck in ionic channel (upper panel of (c-f)) and electronic channel (lower panel of (c-f)). As in **Fig. 5**, nonlinear base-line $\mu_{Li}$ in (e-f) are simplified to straight lines between boundary potentials, and $\mu_{Li}$ jumps at bottlenecks simplified to zig-zags. If these zig-zags exceed boundary values, their peaks are circled and termed overshoot/undershoot. Note that lengths of $J_{Li^+}$ and (grossly exaggerated) $J_e$

arrows are not accurate and meant for schematics only.

**Appendix 5. Length and time scales required to reach local equilibrium for solid electrolytes**

The combined thermodynamic and kinetic analysis in this work relies on the critical though "standard" [22-24, 36] assumption of local equilibrium in the context of irreversible thermodynamics, which requires interacting species to equilibrate at a fine enough length scale and short enough time scale, so that thermodynamic laws still govern their energetics and kinetic laws still govern their transport. Below, we estimate these time and length scales for YSZ and LLZO solid electrolytes under their typical operating conditions.

The time scale for establishing local equilibrium should be smaller than the practical timescale for cell operations, which ranges from hours to days for SOFC/SOEC and batteries. One way to estimate the time scale for reaching local equilibrium is to use the upper and lower limits of stoichiometry under the most oxidizing and most reducing conditions in the operation, and find the time required to transport the species to achieve the stoichiometry change. For YSZ, the oxygen nonstoichiometry $\delta$ of $Y_{0.148}Zr_{0.852}O_{1.926-\delta}$ under the most reducing and most oxidizing conditions in **Fig. 3a** ($\mu_{O_2}$ of −5.21 eV and 0.36 eV, respectively), is obtained from

$$\delta^2(\delta + 0.074)\exp(\frac{\mu_{O_2}}{2k_BT}) = 0.2276\exp(\frac{-3.98 \text{ eV}}{k_BT}) \qquad (A19)$$

according to the thermodynamic data in Ref. [A1]. It gives $\delta$ of 0.001 at $\mu_{O_2} = -5.21$ eV and $3\times10^{-10}$ at $\mu_{O_2} = 0.36$ eV. To fully alter the stoichiometry of an entire 10 μm-thick YSZ membrane over this range (from $3\times10^{-10}$ to 0.001, which is an over-estimate) under an operational current (mostly ionic) of $-1$ A/cm$^2$ requires a time of

$$\frac{2\times(\text{stoichiometry change})\times(\text{thickness})\times\dfrac{(\text{density})}{(\text{molecular weight})}\times(\text{Faraday constant})}{\text{current density}}$$

$$=\frac{2\times(0.001-3\times10^{-10})\times(10\text{ μm})\times\dfrac{6.10\text{ g/cm}^3}{121.7\text{ g/mol}}\times(96485\text{ C/mol})}{1\text{ A/cm}^2}=9.6\times10^{-3}\text{ s}$$

(A20)

A similar calculation can be performed for LLZO. To achieve a Li stoichiometry change of 0.001 of a 10 μm-thick LLZO membrane at an operational current density (mostly ionic) of $-0.1$ mA/cm$^2$ requires

$$\frac{(\text{stoichiometry change})\times(\text{thickness})\times\dfrac{(\text{density})}{(\text{molecular weight})}\times(\text{Faraday constant})}{\text{current density}}$$

$$=\frac{0.001\times(10\text{ μm})\times\dfrac{5.2\text{ g/cm}^3}{839.7\text{ g/mol}}\times(96485\text{ C/mol})}{0.1\text{ mA/cm}^2}=6.0\text{ s}$$

(A21)

Note in Eq. (A20-A21), we assume all ionic current is to alter the stoichiometry but not to anticipate electrode reactions, so the estimates should be treated as a lower bound of the equilibrium time. Nevertheless, the much shorter timescale obtained in the above than the practical one for cell operation suggest the local equilibrium is likely to be reach rapidly.

The length scale over which local equilibrium is established should be much smaller than the size of electrochemical cells and the length scale of the critical microstructure for thermodynamic and kinetic analysis. The shortest of these latter length scales are in the range of tens of μm (such as cell thickness). One way to estimate the length scale to reach local equilibrium is to find the spacing between reacting species, for it is this distance that they must travel before they can react. Since solid electrolytes used for electrochemical cells are all fast ion conductors, ions can travel freely to where an electron/hole is located, and to react, so it is the average spacing between electrons (or holes) that sets the diffusion distance. For YSZ, Park and Blumenthal gave the electron/hole density of $10^{16\text{-}17}/cm^3$ at 800-1000 $^oC$ in Ref. [28]. Therefore, every cube of 20-50 nm in size should have one electron (or hole). This is the average distance that an $O^{2-}$ (or oxygen vacancy) will random-walk to meet an electron/hole and to react with it to establish local equilibrium. Such distance is short compared to the cell dimensions and the length scale of critical microstructure in YSZ cells.

The diffusion distance estimated above provides us another method to estimate the time required for establishing equilibrium reaction. This is the diffusion time for the $O^{2-}$ (or oxygen vacancy) to reach an electron/hole. Using

$$t = \frac{(\text{diffusion distance})^2}{6 \times \text{diffusivity}} \quad (A22)$$

for 3-dimensional random walk with a diffusion distance of 20-50 nm and oxygen diffusivity of $10^{-8}$-$10^{-7}$ $cm^2/s$ at 800 $^oC$ [A2, A3], we obtain a time of $7 \times 10^{-6}$ s - $4 \times 10^{-4}$ s. Once again, the time is short compared to the typical operational time of a YSZ cell.

**References**

A1. J.H. Park, R.N. Blumenthal, Thermodynamic properties of nonstoichiometric

yttria-stabilized zirconia at low oxygen pressures, J. Am. Ceram. Soc., 72 (1989) 1485-1487.

A2. P.S. Manning, J.D. Sirman, R.A. De Souza, J.A. Kilner, The kinetics of oxygen transport in 9.5 mol% single crystal yttria stabilised zirconia, Solid State Ionics, 100 (1997) 1-10.

A3. M. Kilo, C. Argirusis, G. Borchardt, R.A. Jackson, Oxygen diffusion in yttria stabilised zirconia—experimental results and molecular dynamics calculations, Phys. Chem. Chem. Phys., 5 (2003) 2219-2224.

**Appendix 6. Time required for electrons to communicate with an internal precipitate**

Because the internal precipitates ($O_2$ bubbles, voids, or Li or Na metal islands) are all neutral, whereas the ionic species that aggregate to form them ($O^{2-}$, oxygen vacancies, $Li^+$, and $Na^+$) are charged, the excess charge must be removed by electron or hole transport across the solid electrolyte. If the time required to transport electrons or holes is short compared to the normal parameters of cell operation, then precipitate formation is entirely determined by thermodynamics (i.e., whether the chemical potential of $O_2$ or Li/Na is high enough) as assumed in our work. Otherwise, the precipitates may not form because additional thermodynamic considerations are needed (a charged entity is energetically costly), or because it is kinetically inhibited by charge transport. Using the conductivity data in **Fig. 1a** at 800 °C for YSZ and an electronic transference number of 0.0022 for **Fig. 3a**, and under an operational (mostly ionic) current density of −1 A/cm², we estimate the time required for electron transport in YSZ to neutralize a precipitating oxygen bubble of 100 nm size with 10 atm internal oxygen pressure (hence ideal gas law still valid, with atm converted to Pa), over a 100×100 μm² area that the bubble covers, is

$$\frac{4\times\dfrac{(\text{volume})\times(\text{pressure})}{(\text{gas constant})\times(\text{temperture})}\times(\text{Faraday constant})}{(\text{area})\times(\text{current density})\times(\text{electronic transference number})}$$

$$=\frac{4\times\dfrac{4}{3}\pi\times(50\text{ nm})^3\times(10\text{ atm})}{(8.314\text{ J}\cdot\text{K}^{-1}\cdot\text{mol}^{-1})\times(1073\text{ K})}\times(96485\text{ C/mol})}{(100^2\text{ μm}^2)\times(1\text{ A/cm}^2\times 0.0022)} = 1.0\times10^{-7}\text{ s} \quad (A23)$$

Likewise, in a LZZO with a Li$^+$ conductivity of ~$10^{-4}$ S/cm and an electronic conductivity ~$10^{-8}$ S/cm (thus an electronic transference number of $10^{-4}$), under an operational current density of 0.1 mA/cm², the time required for electron transport to neutralize a 100 nm Li metal (island) precipitate covering 100×100 μm² area of LLZO is

$$\frac{(\text{volume})\times\dfrac{(\text{density})}{(\text{molecular weight})}\times(\text{Faraday constant})}{(\text{area})\times(\text{current density})\times(\text{electronic transference number})}$$

$$=\frac{\dfrac{4}{3}\pi\times(50\text{ nm})^3\times\dfrac{0.534\text{ g/cm}^3}{6.941\text{ g/mol}}\times(96485\text{ C/mol})}{(100^2\text{ μm}^2)\times(0.1\text{ mA/cm}^2\times 10^{-4})} = 3.9\text{ s} \quad (A24)$$

Therefore, the time required to deliver electrons/holes to the site of heterophase formation to compensate the charge left by the participating ions is small compared to the operational time of a normal cell.

The same consideration also dictates that this additional current will not disturb the overall current of the cell, i.e., it will not increase the apparent device resistance.

**Appendix 7. Comparison with work of Yoo *et al*.** [40, 41]

Internal phase precipitation depicted in **Fig. 2** cannot be reproduced at blocking interfaces of opposite flux divergence as described by Yoo and Kim [40]. In their experiments, a source-sink pair was formed to support long-range mass transport between them in the same way as in diffusional creep, thereby causing sintering and grain growth at the blocking interfaces. The process is very slow because the diffusion distance is macroscopically long, so it can only be observed under a very large current density and a very high temperature (Yoo and Kim used 15-25 A/cm$^2$ and 1100-1200 $^o$C for a relatively nonrefractory CoO). In contrast, at the transverse grain boundaries in **Fig. 2**, ion flows are continuous—there is no flux divergence, and the internal phase precipitated contains only a neutral molecular or a metallic species that is not transported in the electrolyte at all—so there is no need for long-range diffusion. As a result, such phase formation, which is a local thermodynamic reaction, is observable under normal operational conditions of SOEC and rechargeable batteries as documented in **Table 1**.

In another paper, Yoo *et al*. [41] described phase decomposition under an electric potential beyond the stability window of a solid electrolyte (Bi$_{1.46}$Y$_{0.54}$O$_3$). This concerns a thermodynamic limit, so the instability manifest as the decomposition of Bi$_{1.46}$Y$_{0.54}$O$_3$ should occur regardless of the microstructure, for example, whether Bi$_{1.46}$Y$_{0.54}$O$_3$ is a single crystal or not. In contrast, phase precipitation depicted in **Fig. 2** and the phase dissolution described in the main text arise from the potential overshoot/undershoot at grain boundaries, so the electrolyte is stable if it is a single crystal. This is shown in **Fig. 1b**, which assumes a single crystal: $\mu_{O_2}$ is bounded by the boundary potentials at $x=0$ and $L$, so it should not cause any unexpected phase to form at all.